\documentclass[aps,prb,twocolumn]{revtex4}
\usepackage{graphicx}
\begin{document}
\title
{On the evaluation of matrix elements in partially projected wave functions}
\author{Noboru Fukushima$^1$, Bernhard Edegger$^1$, 
		  V. N. Muthukumar$^2$, Claudius Gros$^1$ }
\affiliation{$^1$
Department of Physics, University of the Saarland,
66041 Saarbr\"ucken, Germany}
\affiliation{$^2$ 
Department of Physics, City College at the City University of New York,
New York, NY 10031
}
\date{\today}

\begin{abstract}
We generalize the Gutzwiller approximation scheme
to the calculation of nontrivial matrix elements between
the ground state and excited states. In our scheme,
the normalization of the Gutzwiller wave function
relative to a partially projected wave function with
a single non projected site (the reservoir site)
plays a key role. For the Gutzwiller projected Fermi sea,
we evaluate the relative normalization both analytically
and by variational Monte-Carlo (VMC). We also report VMC
results for projected superconducting states that show 
novel oscillations in the hole density near the reservoir site.
\end{abstract}
\maketitle

\section{Introduction}
This paper concerns the calculation of matrix elements using 
projected wave functions of the form, 
$|\Psi \rangle = P~|\Psi_0 \rangle$. Here,
$P = \prod_i (1-n_{i\uparrow} n_{i\downarrow})$ 
is a projection operator which
excludes double occupancies at sites $i$, and 
$|\Psi_0 \rangle$,
a trial wave function.
Projected wave functions of this form 
were originally
proposed by Gutzwiller to study electronic systems with
repulsive on-site interactions \cite{gutzwiller_63}. 
The choice of $|\Psi_0 \rangle$ depends on the problem under
consideration. For instance, a projected Fermi liquid state,
\begin{equation}
P\ |\Psi_{{\rm FS}}\rangle\ =\ P\ \prod_{k<k_F}
 c_{k\uparrow}^\dagger c_{k\downarrow}^\dagger |0\rangle~,
\label{psi0_FS}
\end{equation}
was used successfully in the
description of liquid $^3$He as an almost localized
Fermi liquid \cite{seiler_86, gros87_almost}.
Soon after the discovery of high temperature 
superconductivity in the cuprates,
projected BCS wave functions
were proposed as possible ground states
of the so-called $t-J$ model 
\cite{pwa_87, bza_87}. 
Early results from
variational Monte Carlo (VMC) studies as well as a renormalized mean 
field theory
based on Gutzwiller approximation showed that 
a projected $d$-wave BCS state,
\begin{equation}
P_N P \ |\Psi_{{\rm BCS}}\rangle= P_N P \ \prod_{k}\left(
u_k+v_k c_{k\uparrow}^\dagger c_{-k\downarrow}^\dagger
\right)|0\rangle~,
\label{psi0_BCS}
\end{equation}
reproduces many features seen in the phase diagram of the 
high temperature superconductors \cite{gros_88, RMFT, atoz}.
The projection operator $P_N$ which fixes the particle number
$N$ in (\ref{psi0_BCS}), is useful when considering
the phase diagram near half filling \cite{gros_88}.
Without $P_N$ in (\ref{psi0_BCS}), one would need to
consider the effects of particle number fluctuations,
which become singular near half-filling \cite{pwa_tunnel,yokoyama88}.

Detailed VMC studies have been carried out recently using 
projected $d$-wave BCS states as
variational wave functions for the two dimensional
Hubbard model \cite{paramekanti}, after a suitable
canonical transformation \cite{gros87_almost}. Similar wave functions
have been proposed in the literature for cobaltate superconductors
as well as organic superconductors \cite{ogata_03,wang_04}. 
To make analytical progress however, 
it is desirable to extend Gutzwiller's scheme and
construct normalized single particle excitations and
calculate matrix elements.
In this paper, we take the first step in this direction. 
We construct normalized excitations of the Gutzwiller projected 
Fermi sea and consider the evaluation of matrix elements.

In his original paper, Gutzwiller proposed that 
in calculating expectation values of operators with projected
wave functions, the effects 
of projection on the state $|\Psi_0 \rangle$ could be 
approximated by a classical statistical weight factor, which
multiplies the quantum result \cite{vollhardt_84}. Thus, for example,
\begin{equation}
{\langle \Psi| \hat{O} |\Psi\rangle\over\langle \Psi|\Psi\rangle}\ \approx \ g\,
{\langle \Psi_0| \hat{O} |\Psi_0\rangle\over\langle \Psi_0|\Psi_0\rangle}~,
\label{renorm}
\end{equation}
where $\hat{O}$ is any operator, and $g$, a statistical weight
factor. 
The basic idea is that the projection operator $P$ 
reduces the number of allowed states in the Hilbert space,
and invoking a simple approximation, such a reduction can be
taken into account through combinatorial factors.
For example, expectation values of the 
the kinetic energy operator 
$c^\dagger_i c_j +c^\dagger_j c_i$ 
and the superexchange interaction between sites
$i$ and $j$, ${\vec S}_i \cdot {\vec S}_j$ in the projected
subspace of states are renormalized by the Gutzwiller
factors,
\begin{equation}
g_t={1-n\over 1-n/2}, \quad
g_s={1\over (1-n/2)^2}~,
\label{g_t_s}
\end{equation}
where $n$ is the density of electrons.
In deriving these renormalization factors,
one considers
the number of states that contribute to
$ \langle \Psi| \hat{O} |\Psi\rangle $ and to
$\langle \Psi_0| \hat{O} |\Psi_0\rangle $ respectively.
The ratio of these two contributions is
identified as the renormalization factor.

It is clear that this approach can be 
generalized to evaluate matrix elements of an operator $\hat{O}$
between different projected states. However, as we will see in this paper, 
many of the matrix elements that are of interest 
are reduced to the calculation of matrix elements between
partially projected wave functions of the form
\begin{equation}
 |\Psi_l'\rangle \ =\ P_l'\ |\Psi_0\rangle , \qquad
 P_l' = \prod_{i\ne l}(1-n_{i\uparrow} n_{i\downarrow})~.
\label{Psi_l}
\end{equation}
The wave function $|\Psi_l'\rangle$ describes a state where
double occupancies are projected out on all sites
except the site $l$, which we call the reservoir site.
The reason for the appearance of reservoir sites
is not far to seek. Consider, for
example, the operator $Pc_{l\uparrow}$. Clearly, it can be 
rewritten as 
$c_{l\uparrow} P_l'$. 
Since calculation of matrix elements involving excited
states involve the commutation
of projection operators with creation/destruction operators,
partially projected states arise inevitably
within the Gutzwiller scheme.

In this paper, we present a method to calculate matrix elements
between a partially projected Fermi sea, {\em i.e.}, a projected 
Fermi sea with a reservoir site at $l$, as in (\ref{Psi_l}). We
will show that this problem has to be solved 
if we were to construct
normalized particle/hole excitations of the (fully) projected Fermi sea.
The same problem arises when calculating
matrix elements for particle/hole tunneling into the projected Fermi sea.
We develop an analytical approximation to solve this problem, and
use it to calculate various matrix elements. We use VMC to test the
validity of the approximation and find that our analytical results
for the partially projected Fermi sea are in good agreement with the 
results from VMC.

The outline of the paper is as follows. In 
Sec.\ \ref{sect_occupancy}, we present results
for the occupancy of the reservoir site. 
We use these results in Sec.\ \ref{sect_excitations},
where we show how normalized single particle excitations 
can be constructed from the projected Fermi sea. In 
Sec.\ \ref{sect_tunneling} we calculate the
matrix elements for particle/hole tunneling into the 
projected Fermi sea. VMC results for 
density oscillations in the vicinity of the
reservoir site for both projected Fermi sea and BCS states
are presented in Sec.\ \ref{sect_oscillations}.
The final section contains
a summary and discussion of results.
\begin{figure}
\medskip
\includegraphics[width=7.5cm,keepaspectratio]{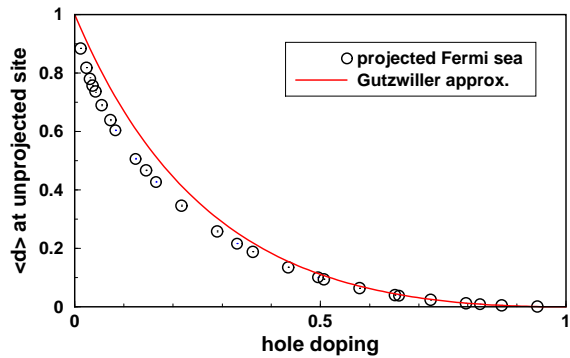}
\caption{Double occupancy of the reservoir site,
$\langle n_\uparrow n_\downarrow\rangle _{\Psi_l'}=1-X$,
as a function of doping,
for the partially projected
Fermi sea. See (\ref{psi_l}) and (\ref{psi0_FS}).
Note the good agreement between the
Gutzwiller result (solid line), Eqs.\
(\ref{X_Gutzwiller},\ref{psi_l_double}) and the VMC results
for the projected Fermi sea (open circles).
Statistical errors and finite-size corrections are estimated
to be smaller than the symbols.
}
\label{d_Gutz_doping}
\end{figure}

\section{Occupancy of the reservoir site}\label{sect_occupancy}

Consider a partially projected wave function,
\begin{equation}
 |\Psi_l'\rangle \ =\ P_l'|\Psi_0\rangle , \qquad
 P_l' = \prod_{i\ne l}(1-n_{i\uparrow} n_{i\downarrow})~.
\label{psi_l}
\end{equation}
Double occupancy is projected out on all sites
except the site $l$, called the reservoir site.
Unless specified otherwise, we take $|\Psi_0\rangle$ to mean the
Fermi sea.
For the calculation of single particle excitations
and matrix elements, we need
expectation values such as
\begin{equation}
{\langle \Psi_l'| \hat{O} |\Psi_l'\rangle \over\langle \Psi_l'|\Psi_l'\rangle }\ =\ g'\,
{\langle \Psi_0| \hat{O} |\Psi_0\rangle \over\langle \Psi_0|\Psi_0\rangle }~, 
\label{renorm_l}
\end{equation}
that generalize the Gutzwiller renormalization
scheme (\ref{renorm}) to partially projected wave functions.

\subsection{Gutzwiller approximation}

In order to evaluate the generalized renormalization 
parameters $g'$ in (\ref{renorm_l}), we obviously need
the normalization $\langle \Psi_l'|\Psi_l'\rangle $.
We define
\begin{equation}
 X\ =\ {\langle \Psi_0|P P |\Psi_0\rangle \over \langle \Psi_0|P_l' P_l'
  |\Psi_0\rangle }\ = 
  {\langle \Psi |\Psi \rangle \over \langle \Psi_l'|\Psi_l'\rangle}~,
\label{def_X}
\end{equation}
the norm of the fully projected state relative to the state
with one reservoir site. Invoking
the Gutzwiller approximation, 
we estimate this ratio by considering
the relative sizes of the Hilbert spaces,
\begin{equation}
X\ \sim  { {L!\over N_\uparrow!N_\downarrow! N_h!}
\over
{L!\over N_\uparrow!N_\downarrow! N_h!} + 
{(L-1)!\over (N_\uparrow-1)!(N_\downarrow-1)! (N_h+1)!} 
}~,
\label{cal_X}
\end{equation}
where $L=N_\uparrow+N_\downarrow+N_h$, is the number of lattice sites,
$N_\uparrow$, $N_\downarrow$ and $N_h$, the number of up spins,
down spins and empty sites respectively. The first term
in the denominator of (\ref{cal_X}) represents the number of
states with the reservoir site being empty or singly occupied;
the second term represents the state with the reservoir site
being doubly occupied.

Eq. (\ref{cal_X}) can be simplified in the thermodynamic limit. We get,
\begin{equation}
X\ = \ {1-n\over (1-n_\uparrow)(1-n_\downarrow)}~,
\label{X_Gutzwiller}
\end{equation}
where the particle densities,
$n_\sigma = N_\sigma/L$ ($\sigma=\ \uparrow,\downarrow$)
and $n=n_\uparrow+n_\downarrow$.
The above argument can be extended to the case of two unprojected
sites in an otherwise projected Fermi sea. We then get,
\begin{equation}
 {\langle \Psi_0|P P |\Psi_0\rangle \over \langle \Psi_0|P_{lm}' P_{lm}'
  |\Psi_0\rangle }\ =\ X^2 \ ,
\label{calc_X2}
\end{equation}
where, $P_{lm} = \prod_{i\ne l,m}(1-n_{i,\uparrow}n_{i,\downarrow})$.
We note for later use that
\begin{equation}
{1-X\over X}\ = \ 
{n_\uparrow n_\downarrow\over (1-n)}~.
\label{(1-X)/X}
\end{equation}

\subsection{Exact relations}

Assuming translation invariance,
it is possible to derive the following exact expressions
\begin{eqnarray} \label{psi_l_empty}
 \langle (1-n_{l\uparrow})(1-n_{l\downarrow})\rangle _{\Psi_l'} & =& X(1-n)\\
\label{psi_l_single}
 \langle n_{l\sigma}(1-n_{l-\sigma})\rangle _{\Psi_l'} & =& X n_\sigma \\
\label{psi_l_double}
 \langle d\rangle _{\Psi_l'} \ \equiv\  
 \langle n_{l\uparrow}n_{l\downarrow}\rangle _{\Psi_l'} & =& 1-X
\end{eqnarray}
for the occupancy of the reservoir site, where 
$$
\langle ... \rangle _{\Psi_l'}\equiv
\langle \Psi_l'|  ... |\Psi_l' \rangle / \langle \Psi_l'|\Psi_l'
\rangle~.
$$
The proof is 
straightforward. Consider for instance, the probability (\ref{psi_l_empty}) 
of finding
the reservoir site empty. Since,
\begin{eqnarray}
&&\langle \Psi_0|P (1-n_l) P|\Psi_0\rangle  
 \qquad \qquad \label{note_that} \\ \nonumber
 && \qquad \qquad
\ =\ \langle \Psi_0|P_l' (1-n_{l\uparrow})(1-n_{l\downarrow}) P_l'|\Psi_0\rangle 
\end{eqnarray}
we have,
\begin{eqnarray*}
&& \langle (1-n_{l\uparrow})(1-n_{l\downarrow})\rangle _{\Psi_l'} 
 \qquad \qquad \\ 
 && \qquad
\ =\ 
{\langle \Psi|(1-n_l)|\Psi\rangle  \over \langle \Psi|\Psi\rangle  }
{\langle \Psi|\Psi\rangle  \over \langle \Psi_l'|\Psi_l'\rangle  }
\ =\ (1-n)X~.
\end{eqnarray*}
Eqs.\ (\ref{psi_l_single}) and (\ref{psi_l_double}) 
can be proved analogously. 

\subsection{VMC results for projected Fermi sea and BCS states}

\begin{figure}
\medskip
\includegraphics[width=7.5cm,keepaspectratio]{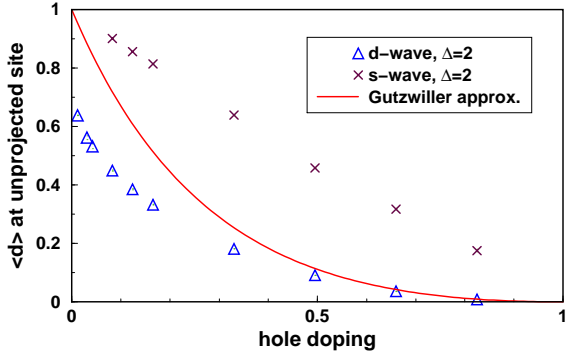}
\caption{Double occupancy at the reservoir site
$\langle n_\uparrow n_\downarrow\rangle _{\Psi_l'}=1-X$,
as a function of doping,
for the partially projected
BCS wave function, see (\ref{psi_l}) and (\ref{psi0_BCS}).
The parameterization follows Ref.\ \onlinecite{gros89_annals}.  
Statistical errors and finite-size corrections are estimated
to be smaller than the symbols.
}
\label{d_BCS_doping}
\end{figure}

In Fig.\  \ref{d_Gutz_doping}, we compare (\ref{X_Gutzwiller})
with VMC results for $\langle d\rangle _{\Psi_l'}=1-X$. 
We find that the results from the generalized
Gutzwiller approximation are in excellent qualitative agreement with the
VMC results for a partially projected Fermi sea.
We also used VMC to obtain the same quantity using projected
$s$/$d$-wave BCS states as variational states in the simulation.
The results for $\langle d\rangle _{\Psi_l'}$
in BCS states are shown in Fig.\  \ref{d_BCS_doping}.
In contrast to the projected Fermi sea, a clear deviation
from the Gutzwiller approximation is seen. This underscores
the importance of pairing correlations in the unprojected
wave function that are not completely taken into
account by the Gutzwiller approximation scheme.
These differences between Fermi sea and BCS states are discussed in
more detail in Sect.\ \ref{sect_oscillations}, where we consider
density oscillations in the vicinity of the reservoir site.

In the following we discuss some
details of the VMC calculations with one unprojected (reservoir) site $l$.
As mentioned earlier, single occupancy is enforced (by projection) on
all other sites.
Simulations are performed on a finite square lattice spanned by two
vectors $(L_x,L_y)$ and $(-L_y,L_x)$ with periodic boundary
conditions \cite{gros89_annals}.
The number of sites, $L=L_x^2+L_y^2$. The numbers of up- and
down-electrons are chosen to be equal, 
${N_{\uparrow}=N_{\downarrow}}$.
The simulation for the local quantity 
$\langle d\rangle _{\Psi_l'}=\langle n_{l\uparrow}
n_{l\downarrow} \rangle_{\Psi_l'}$ has a larger statistical error than 
results for macroscopic quantities in uniform systems 
because the summation over site indices yields effectively $L$ times more
statistics for the latter.
In order to overcome this problem, we update the
reservoir site more often than the projected sites.
Accordingly, the transition probability needs an extra 
weighting factor to keep the local balance. With this 
procedure, we can improve the statistical accuracy by
about one order of magnitude.
In addition, we carry out measurements after every update. Usually, in
VMC simulations, measurement are performed every $O(L)$ updates to obtain
independent samples since similar states return similar sampled data.
However, in the case of $n_{l\uparrow} n_{l\downarrow}$, a measurement
returns only 0 or 1; \textit{viz.}, the sampled data
can be different even when the states are similar.
Given this, the measurement after every update seems more 
reasonable as it reduces statistical errors.
Furthermore, we have restricted updates to the transfer of a 
single electron to an unoccupied site, 
and excluded updates \textit{via} the exchange of two electrons. 
The calculation of the transition probabilities for the former 
update consumes time of $O(N_\sigma)$
whereas the time taken for the latter update is $O(N_\sigma^2)$.
As the system size increases, this
restriction achieves efficiency.
We have collected statistics from up to 60 independent runs over
two days, and the total number of updates 
amounts to $10^8 \sim 10^9$.

For superconducting states, one can perform the VMC simulation 
either with fixed
particle number $P_N P |\Psi_{\rm BCS}\rangle$, or with a
fixed phase $P |\Psi_{\rm BCS}\rangle$ \cite{gros89_annals,yokoyama88}. 
For the latter choice, particle number fluctuation hinders 
the variational wave function from reaching half filling 
unless the chemical potential $\mu$ goes to
infinity.  On the other hand, the wave function
can be optimized by varying the gap $\Delta_k$ even at 
half filling, if we choose to fix the particle number.
It is important to note that
simulations with fixed particle number are done not with the
most probable $N$ of $P|\Psi_{\rm BCS}\rangle$, but that of
$|\Psi_{\rm BCS}\rangle$. This is because $P$ decreases the average particle
number \cite{yokoyama88}. Despite these differences,
both choices of wave functions yield quantitatively
similar results \cite{yokoyama88}. Throughout this paper we choose to fix
the particle number while working with projected BCS states.

Let us define $a_k \equiv v_k/u_k$. For the $d$-wave BCS state, 
$a_{k=0}=0$ in the thermodynamic limit.  
However, if one chooses $a_{k=0}=0$ in the
finite system, the $k=0$ state is unoccupied, although it is the
lowest energy state. One can also choose a large value for $a_{k=0}$.
Usually, the difference between these choices is $O(1/N)$. 
We expect an $N$ electron system with
with $a_{k=0}=0$ to be similar to an $N+2$ electron system
with large $a_{k=0}$, because the two electrons that are no longer in
the $k=0$ state can occupy other available states.  
However, this argument fails at half filling 
for projected states, because 
there are no available states left for
the two extra electrons. So it should not be
surprising that $X$ depends strongly
on these choices close to half filling; the $a_{k=0}=0$ 
definition gives
larger $X$ than the other does as shown in
Fig.~\ref{size-dependence}.  
At the other fillings, 
our results show only $O(1/N)$ of
difference between these choices.
Except for Fig.~\ref{size-dependence}, where we show both
cases, all other results in this paper are 
obtained for a choice of large $a_{k=0}$;
\textit{i.e.}, we take $a_{k=0}$ larger than any other $a_k$.

The system size dependence is quite small except in the vicinity 
of half filling. In fact, it is qualitatively consistent with
the Gutzwiller approximation; 
size dependence enters only as $(N_h+1)/L$ in
Eq.~(\ref{cal_X}) and is negligible for large $N_h$.
In Fig.~\ref{size-dependence}, we show the dependence of
$\langle d \rangle$ on the system size. As shown in Fig.~\ref{size-dependence},
$\langle d \rangle$ approaches unity for the projected Fermi sea.
For the projected $d$-wave BCS state, 
we speculate that the value of $\langle d \rangle$ goes to unity too, 
because it does not saturate, 
but increases more rapidly as $1/L$ decreases.

\section{Single particle excitations of the 
projected Fermi sea} \label{sect_excitations}

We consider the particle excitation
\begin{equation}
|\Psi_{k\sigma}^+\rangle  \,=\, P c_{k\sigma}^\dagger|\Psi_0\rangle~,
\label{psi+}
\end{equation}
and the hole excitation
\begin{equation}
|\Psi_{k\sigma}^-\rangle  \,=\, P c_{k\sigma}|\Psi_0\rangle ~.
\label{psi-}
\end{equation}
Any calculation involving
$|\Psi_{k\sigma}^\pm\rangle $ needs the
respective norms, 
$N_{k\sigma}^\pm=\langle \Psi_{k\sigma}^\pm |\Psi_{k\sigma}^\pm\rangle $.
We now calculate these norms within the generalized Gutzwiller approximation.

\subsection{Particle excitation}

For the particle excitation, we get,
\begin{equation}
{ N_{k\sigma}^+ \over N_G}
= 1-n\, +\, g_{t}\left( n_\sigma - n_{k\sigma}^0 \right)
= g_{t}\left( 1- n_{k\sigma}^0 \right)~ ,
\label{norm_psi+}
\end{equation}
where $g_{t}=(1-n)/(1-n_\sigma)$, $N_G=\langle \Psi|\Psi\rangle$,
and $n_{k\sigma}^0 = \langle 
c_{k\sigma}^\dagger c_{k\sigma}^{\phantom{\dagger}}\rangle _{\Psi_0}$
is the momentum distribution function in the unprojected site.
%

Equation (\ref{norm_psi+}) has appeared frequently
in the literature. here, we repeat its derivation
to facilitate a comparison with
the analogous problem for hole excitations.
The norm $\langle \Psi_{k\sigma}^+|\Psi_{k\sigma}^+\rangle $ is given by
\begin{widetext}

\begin{eqnarray} \nonumber
&&
N_{k\sigma}^+ \ =\ 
\langle \Psi_0|c_{k\sigma}^{\phantom{\dagger}}P Pc_{k\sigma}^\dagger|\Psi_0\rangle  
\ =\ {1\over L}
\sum_{l,m} {\rm e}^{ik(l-m)}
\langle \Psi_0|P_l' (1-n_{l-\sigma})
c_{l\sigma}^{\phantom{\dagger}} c_{m\sigma}^\dagger
(1-n_{m-\sigma}) P_m'|\Psi_0\rangle 
\qquad \\ && \qquad \nonumber
\ =\ {1\over L} \sum_{l} 
\langle \Psi_0|P_l' (1-n_{l\sigma}) (1-n_{l-\sigma}) P_l'|\Psi_0\rangle 
\, + \, {1\over L}
\sum_{l\ne m} {\rm e}^{ik(l-m)}
\langle \Psi_0|P c_{l\sigma}^{\phantom{\dagger}} c_{m\sigma}^\dagger
P|\Psi_0\rangle 
\qquad \\ && \qquad \label{follows}
\ =\ {N_G} 
{\langle \Psi|(1-n) |\Psi\rangle  \over \langle \Psi|\Psi\rangle }
\, - \, {N_G\over L}
\sum_{l\ne m} {\rm e}^{ik(l-m)}
{
\langle \Psi| c_{m\sigma}^\dagger c_{l\sigma}^{\phantom{\dagger}} |\Psi\rangle 
\over \langle \Psi|\Psi\rangle } ~,
\end{eqnarray}
where we have used (\ref{note_that}) for the diagonal
contribution in the last step. Invoking the
Gutzwiller approximation for the off-diagonal term,
Eq.\ (\ref{norm_psi+}) follows directly from
(\ref{follows}).

\subsection{Hole excitation} \label{subsec_hole_excitation}

The normalization of the hole excitation can be done analogously. We get,
\[
{N_{k\sigma}^- \over N_G} 
\ =\ 
{1 \over N_G L} \sum_{l,m} 
{\rm e}^{ik(l-m)}
\langle \Psi_0|P_l' c_{l\sigma}^\dagger
c_{m\sigma}^{\phantom{\dagger}} P_m'|\Psi_0\rangle 
\ =\ 
{1\over X}\Big[ Xn_\sigma + (1-X)\Big]
\,+ \,
{1 \over N_G L} \sum_{l\ne m} {\rm e}^{ik(l-m)} 
\langle \Psi_0|P_{lm}' c_{l\sigma}^\dagger
c_{m\sigma}^{\phantom{\dagger}} P_{lm}'|\Psi_0\rangle ~, 
\]
where,
$P_{lm} = \prod_{i\ne l,m}(1-n_{i,\uparrow}n_{i,\downarrow})$.
The last term in the above equation corresponds to a hopping
process between two reservoir sites. The
generalized Gutzwiller approximation assumes 
that the matrix elements are proportional to the
square roots of the corresponding densities 
(\ref{psi_l_empty},\ref{psi_l_single},\ref{psi_l_double}).

Invoking the Gutzwiller approximation and using (\ref{calc_X2}),
we get,
\begin{equation} \label{norm_psi-}
{N_{k\sigma}^- \over N_G} \ =\ 
{\langle \Psi_{k\sigma}^- |\Psi_{k\sigma}^-\rangle   \over
\langle \Psi|\Psi\rangle  } \ =\
n_{\sigma} \,+\, {1-X\over X}\, +\,
{n_{k\sigma}^0-n_\sigma \over X^2 n_\sigma(1-n_\sigma)}
\Big[\,
\sqrt{X(1-n)} \sqrt{X n_\sigma}
+ \sqrt{X n_{-\sigma}} \sqrt{1-X}
\,\Big]^2~,  \label{eq_holeex1}
\end{equation}
for the normalization of the hole excitation relative
to the norm of the Gutzwiller wave function.

The general expression (\ref{norm_psi-}) for
the hole normalization, can be simplified upon
using the Gutzwiller result
(\ref{(1-X)/X}) for the relative norm $X$. We then get,
\[
{n_{k\sigma}^0-n_\sigma \over n_\sigma(1-n_\sigma)}
\Big[\,
\sqrt{1-n} \sqrt{ n_\sigma}
+ \sqrt{ n_{-\sigma}} \sqrt{(1-X)/X}
\,\Big]^2
=\ (n_{k\sigma}^0-n_\sigma)
{[ (1-n) + n_{-\sigma} ]^2
\over (1-n_\sigma)(1-n)}
=\ (n_{k\sigma}^0-n_\sigma)
{1-n_{-\sigma} \over (1-n)}~,
\]
for the last term in (\ref{norm_psi-}).
Finally, we get the simple result,
\end{widetext}

\begin{equation} \label{norm_psi-X}
{N_{k,\sigma}^- \over N_G} \ =\ 
n_{k\sigma} {1-n_{-\sigma} \over (1-n)}
 \ =\ { n_{k\sigma}\over g_{t}}
\end{equation}
It is interesting to compare this result for the
normalization of the hole excitation with
the corresponding expression (\ref{norm_psi+})
for the particle excitation. The vanishing of
the latter at half filling could have been
expected. But the divergence of  $N_{k\sigma}^-$ as $n \rightarrow 1$
is surprising. We will return to this point in the next section.

\subsection{Consistency check}

The norm $N_{k\sigma}^+$ has to vanish whenever
$ |\Psi_{k\sigma}^+\rangle  \,=\, P c^\dagger_{k\sigma}|\Psi_0\rangle $
vanishes. For the Fermi sea this is the case
when $k < k_F$, i.e.\ when $n_{k \sigma}^0=1$. 
This physical condition is
obviously fulfilled by (\ref{norm_psi+}).
Similarly, we expect 
$N_{k\sigma}^- $ to vanish for $n_{k \sigma}^0=0$,
which is satisfied by (\ref{norm_psi-X}).
Thus,
the Gutzwiller result (\ref{X_Gutzwiller}) obeys 
the normalization condition for the hole excitation and
the theory is consistent.

\begin{figure}[htb]
\includegraphics*[width=7.5cm,keepaspectratio]{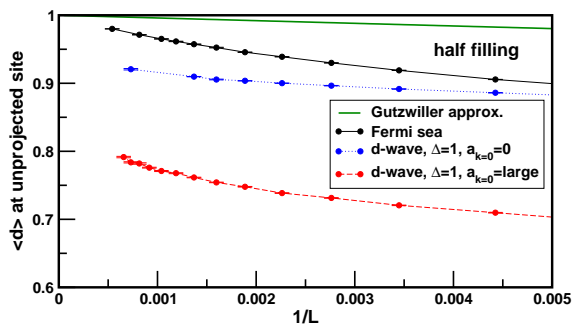}
\caption{The size dependence of the double occupancy at the
reservoir site, (\ref{psi_l_double}), at half filling. 
The VMC result for the Gutzwiller state (upper curve) seems
to converge nicely to unity in the thermodynamic limit,
in agreement with Eqs. (\ref{X_Gutzwiller}) and (\ref{(1-X)/X}).
The results for the projected $d$-wave show a pronounced 
dependence on the occupancy of the $k=0$ state. See text for details.}
\label{size-dependence}
\end{figure}


\section{Tunneling matrix elements}\label{sect_tunneling}

We now consider the tunneling of electrons and holes
into a projected wave function. Single particle
tunneling into
a projected superconducting state has been
considered recently by Anderson and Ong \cite{pwa_tunnel},
and Randeria {\em et al.} \cite{randeria_tunnel}. 
Here, we restrict ourselves
to the projected Fermi liquid state and evaluate
the tunneling matrix elements by retaining
systematically, all terms arising from the commutation
of the electron creation and destruction operators
with the projection operator $P$, as outlined in
Sec.\ \ref{sect_excitations}.

\subsection{Particle tunneling}

Consider first, the matrix element
\begin{equation}
M_{k\sigma}^+ \ =\ 
{ \left| \langle \Psi_{k\sigma}^+| 
 c_{k\sigma}^\dagger |\Psi\rangle  \right|^2
\over N_{k\sigma}^+ N_G }~.
\label{M+}
\end{equation}
The numerator may be calculated easily by using the result of
(\ref{norm_psi+}):
\begin{eqnarray*}
{\langle \Psi_0| c_{k\sigma}^{\phantom{\dagger}} P
 c_{k\sigma}^\dagger P |\Psi_0\rangle   \over N_G}
  &=&  {\langle \Psi_0| c_{k\sigma}^{\phantom{\dagger}} P
  P c_{k\sigma}^\dagger |\Psi_0\rangle   \over N_G} \\
 &= &  g_{t}(1-n_{k\sigma}^0)~. 
\end{eqnarray*}

From the above expression we find that
the particle tunneling matrix element takes the form,
\begin{equation}
M_{k\sigma}^+ \ =\ 
{ g_{t}^2(1-n_{k\sigma}^0)^2
\over g_{t}(1-n_{k\sigma}^0) } \ =\ 
g_{t}(1-n_{k\sigma}^0)~. 
\label{M++}
\end{equation}
It vanishes at half filling $n\to 1$,
implying that
the {\em addition} of electrons is not possible
exactly at half filling because of the restriction in 
the Hilbert space.

\subsection{Hole tunneling}
Next we evaluate the matrix element
\begin{equation}
M_{k\sigma}^- \ =\ 
{ \left| \langle \Psi_{k\sigma}^-| 
 c_{k\sigma}^{\phantom{\dagger}} |\Psi\rangle  \right|^2
\over N_{k\sigma}^- N_G }~,
\label{M-}
\end{equation}
corresponding to the tunneling of holes into the projected state.
Naively, we might expect this process to be allowed at 
half filling, since the {\em removal} of electrons is not
forbidden by the projection operator.
Consider now, the matrix element in the numerator of
(\ref{M-}). 
We follow the same procedure used to evaluate the
norm of the hole wave function in Sec.\ \ref{subsec_hole_excitation}
and use (\ref{(1-X)/X}) and (\ref{psi_l_single}) and find,
\begin{widetext}
\begin{eqnarray} \nonumber &&
{\langle \Psi_0| c_{k\sigma}^\dagger P
c_{k\sigma}^{\phantom{\dagger}} P |\Psi_0\rangle   \over N_G}
 \ =\ 
{1 \over N_G L}\sum_{l,m}
{\rm e}^{ik(l-m)} 
\langle \Psi_0|P_l' c_{l\sigma}^\dagger
c_{m\sigma}^{\phantom{\dagger}} P|\Psi_0\rangle 
\ =\ 
{Xn_\sigma \over X}
\,+ \,
{1 \over N_G L} \sum_{l\ne m} {\rm e}^{ik(l-m)} 
\langle \Psi_0|P_l' c_{l\sigma}^\dagger
c_{m\sigma}^{\phantom{\dagger}} P_l'|\Psi_0\rangle 
\qquad \\ && \qquad
\ =\ 
{Xn_\sigma \over X}
\,+ \, (n_{k\sigma}^0-n_\sigma)
{ \left[\sqrt{Xn_{-\sigma}}\sqrt{1-X}
+\sqrt{X(1-n)}\sqrt{Xn_\sigma} \right]
\sqrt{1-n}\sqrt{n_\sigma}
\over X (1-n_\sigma)n_\sigma }
\label{Following} \\ && \qquad \nonumber  
\ =\
n_\sigma \,+ \, (n_{k\sigma}^0-n_\sigma)
{ n_{-\sigma}n_\sigma
+(1-n)n_\sigma \over  (1-n_\sigma)n_\sigma }
\ =\  n_{k\sigma}^0~.
\end{eqnarray}
\end{widetext}
Using this expression together with
the norm (\ref{norm_psi-}) of the hole excitation,
we obtain the hole tunneling matrix element (\ref{M-}),
\begin{equation}
M_{k\sigma}^- \ =\ 
{ n_{k\sigma}^0n_{k\sigma}^0
\over n_{k\sigma}^0/g_{t}} \ =\
g_{t} n_{k\sigma}^0,
\label{M--}
\end{equation}
a surprising result, in that it vanishes at
half filling (${n_{\uparrow}=n_{\downarrow}=0.5}$) too.

The vanishing of the hole tunneling matrix element at half filling is
clearly related to the divergence of the norm of the hole excitation.
This, in turn, is related to the fact that $X \rightarrow 0$, 
as $n \rightarrow 1$ (\textit{cf.} Eq(\ref{X_Gutzwiller})). 
The vanishing hole tunneling matrix element
can then be understood as follows. 
When the reservoir site is doubly occupied,
a single hole in the otherwise projected Fermi sea can
be found in any of the lattice sites. Consequently, when
double occupancy of the reservoir site occurs with probability 1,
as it does at half filling,
an ``orthogonality catastrophe'' occurs leading to zero
overlap for the tunneling matrix element.
Note that the result (\ref{M--}) hinges on the exact
functional dependence (\ref{(1-X)/X}) of $(1-X)/X$
on the particle densities $n_\sigma$.
On the other hand, the particle tunneling matrix element
$M_{k\sigma}^+$ is not affected
by the functional form of the relative normalization
factor $X$.
If $X$ were to vanish more slowly than $(1-n)$
at half filling, then from
(\ref{norm_psi-}) and (\ref{Following}),
one could conclude that the hole tunneling matrix element
$M_{k\sigma}^-$ does not vanish as $n\to1$, 
possibly leading to an asymmetry
between particle and hole tunneling. Our analytical results
preclude this possibility for the projected Fermi sea. But we are
unable to provide a definite answer for the projected
superconducting states, in view of the discrepancy between the 
Gutzwiller approximation and the VMC results (Fig.\ \ref{d_BCS_doping}).
To understand this discrepancy, we study density oscillations
in the vicinity of the reservoir site using VMC.


\section{Density oscillations near the reservoir site} 
\label{sect_oscillations}

To clarify the limitations of the Gutzwiller
approximation for projected superconducting states, 
we use VMC to calculate the hole density in the vicinity
of the reservoir site. We find that the density oscillations
seen are very different for the projected Fermi sea and the 
BCS states.

\begin{figure*}[htb]
\includegraphics[width=5.5cm,keepaspectratio]{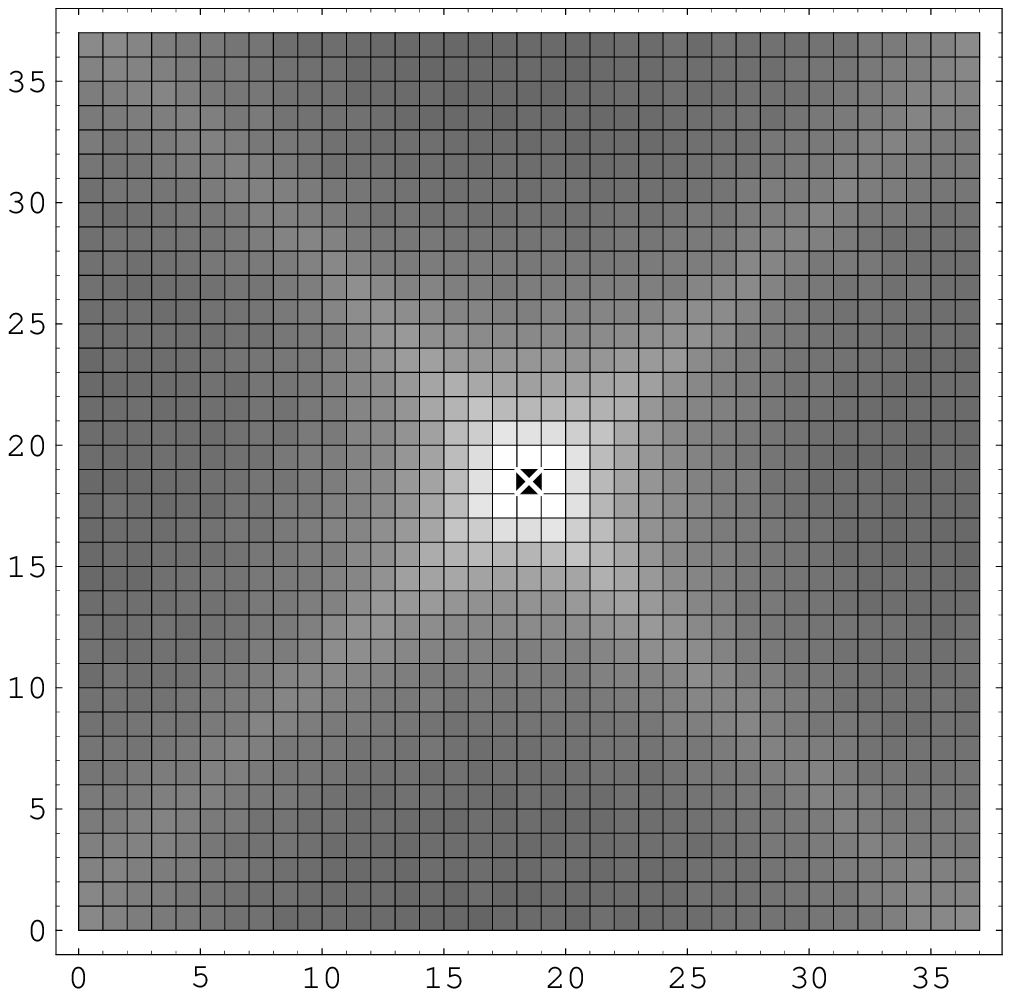}
 \includegraphics[width=5.5cm,keepaspectratio]{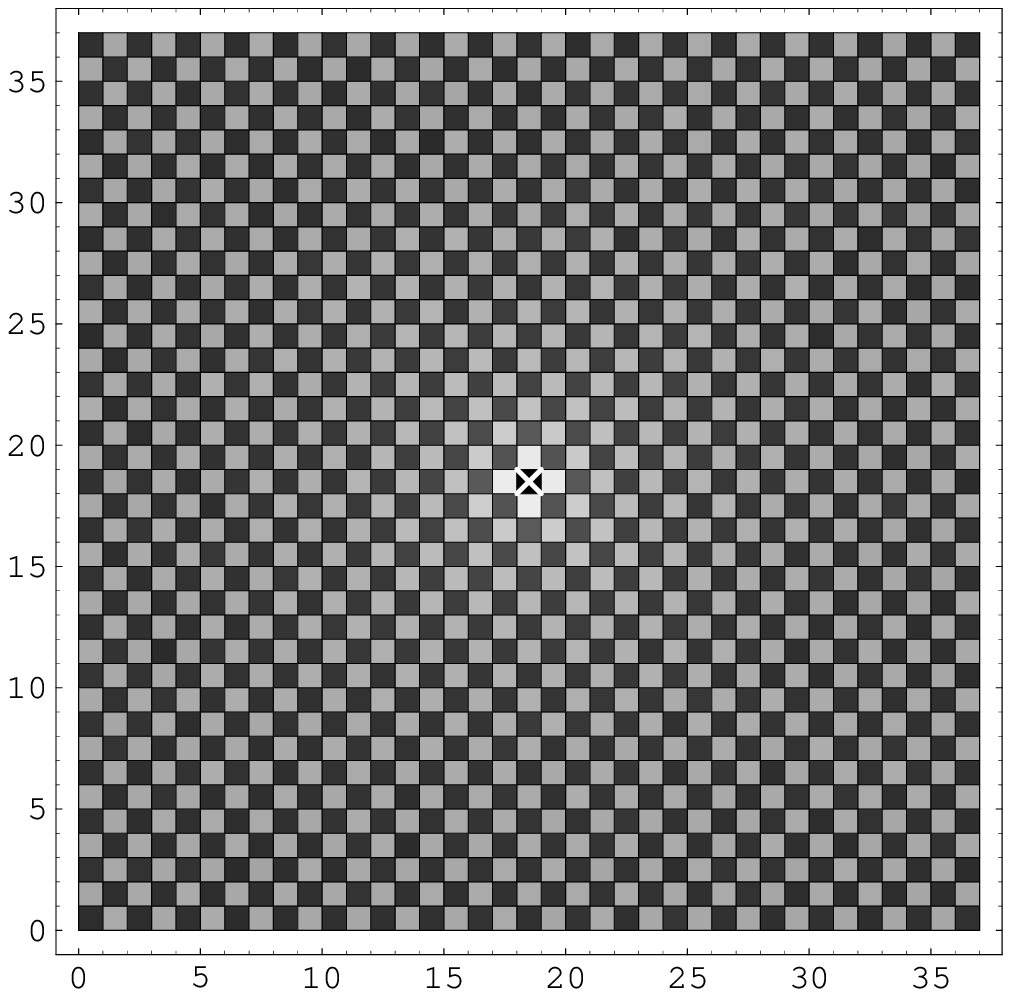}
 \includegraphics[width=5.5cm,keepaspectratio]{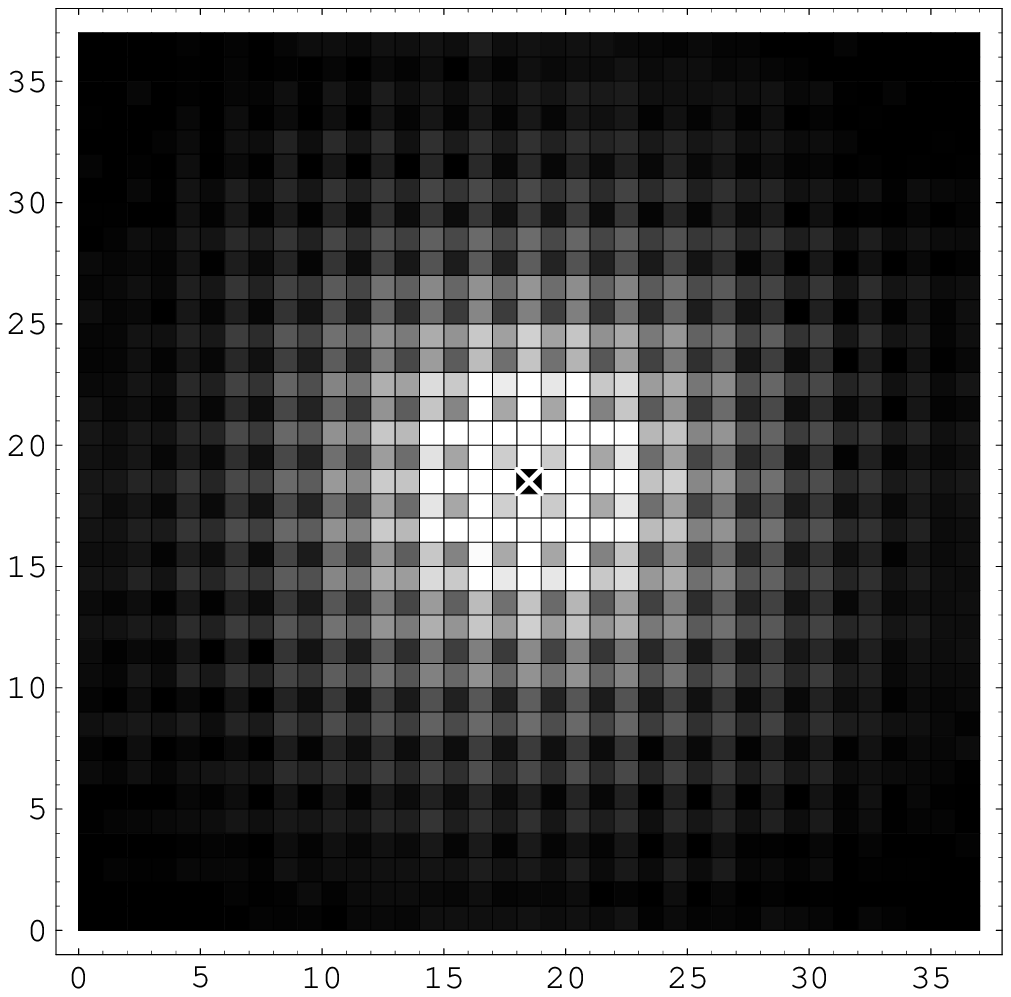}
 \includegraphics[width=5.5cm,keepaspectratio]{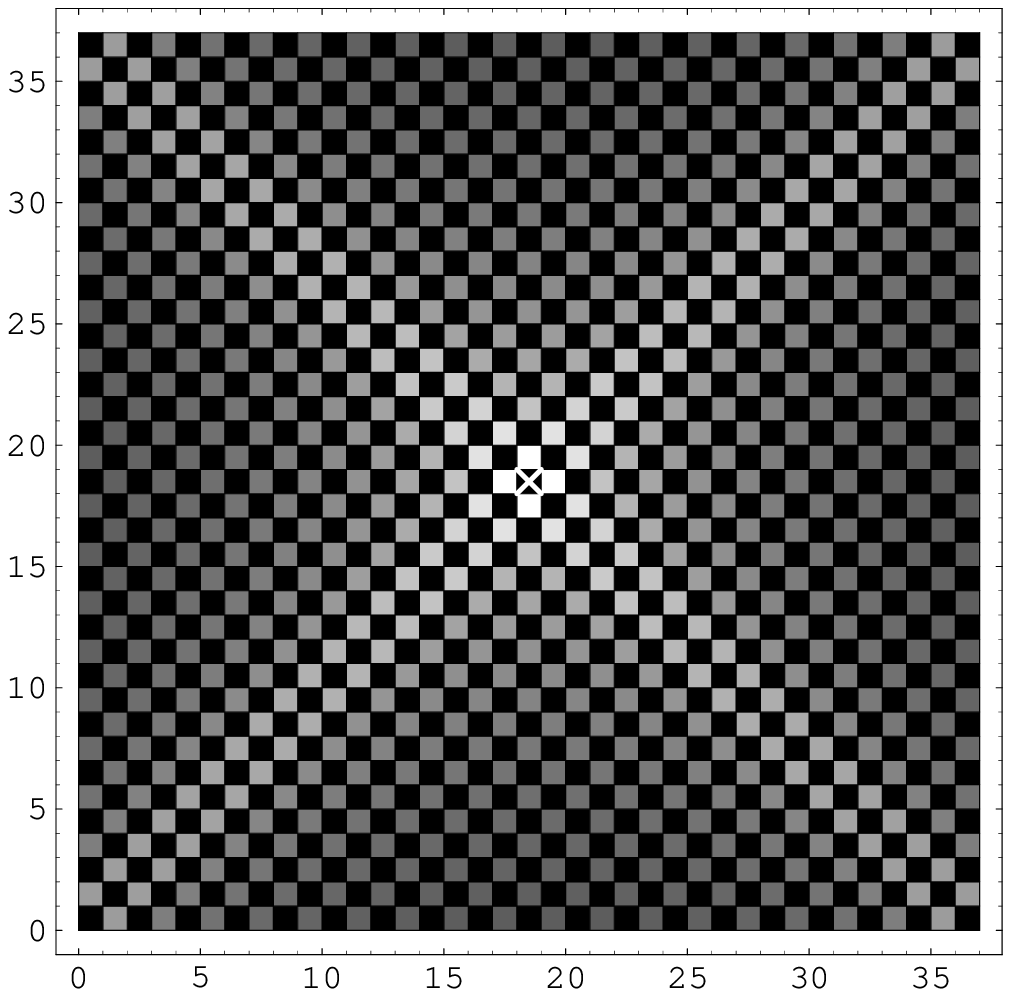}
 \includegraphics[width=5.5cm,keepaspectratio]{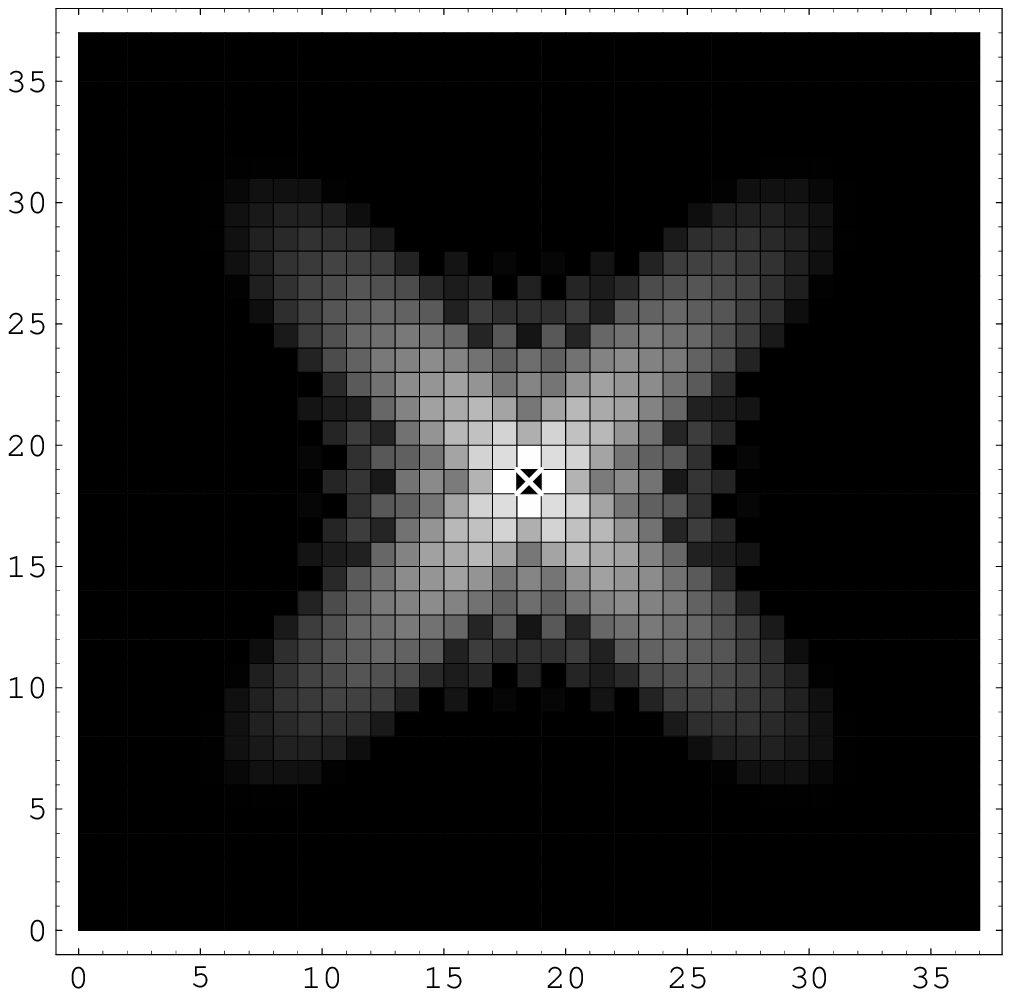}
 \includegraphics[width=5.5cm,keepaspectratio]{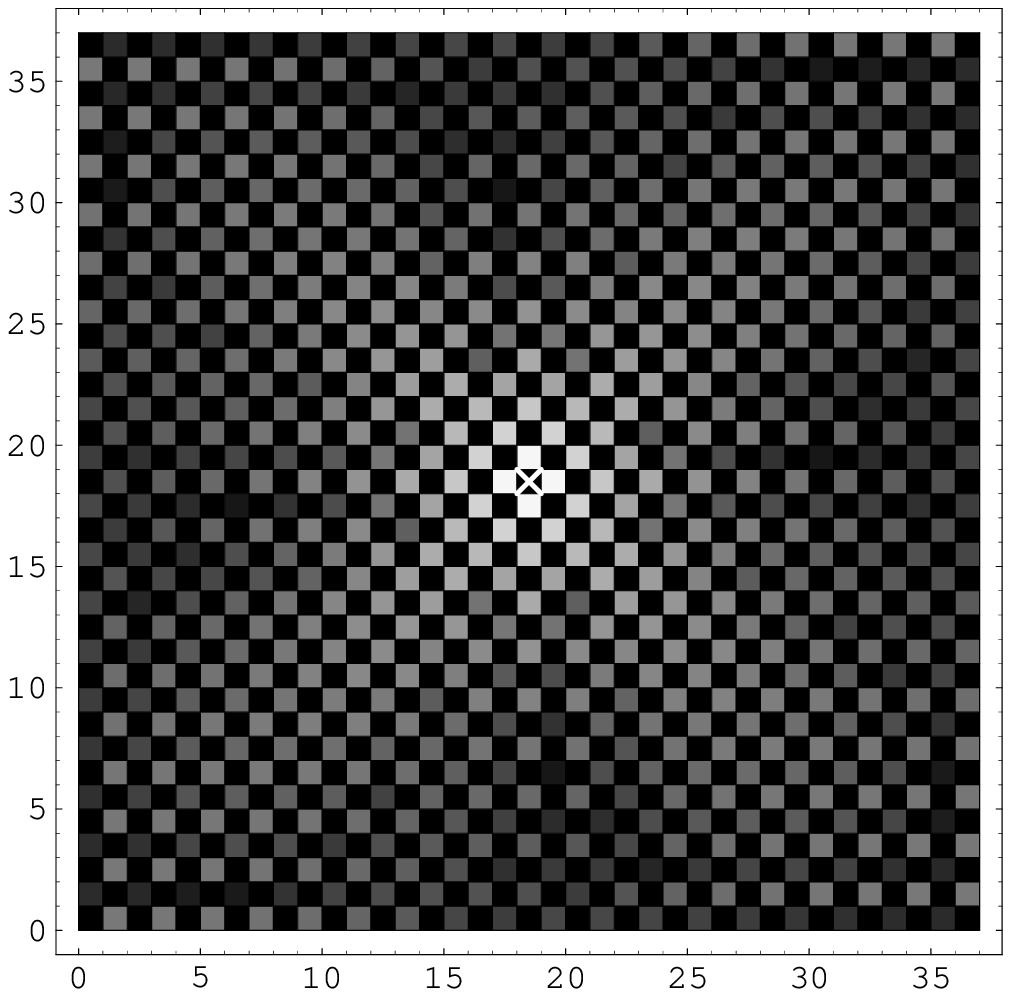}
\caption{{\bf Top row:} VMC results for the hole density
$ n_h(m) = \langle (1-n_m)\rangle _{\Psi_l'}$ 
(color coding: white/black correspond to
high/low values of $n_h(m)$) in the partially projected
state $|\Psi_l'\rangle $, for sites $m$ other than
the reservoir site $l$ (marked by the cross).
Left: Fermi sea. Middle: $s$-wave state. Right: $d$-wave state.
{\bf Second row:} Exact results for
$d_h(m)$, see Eq.\ (\ref{d_h}), in the
unprojected state 
(color coding: white/black correspond to
high/low values), for sites $m$ other than the
doubly occupied site $l$ (marked by the cross).
Left: Fermi sea. Middle: $s$-wave state. Right: $d$-wave state.
{\bf All:} The lattice has $37^2+1=1370$ sites and $n=1$.
For the BCS states, $\Delta=1$.}
\label{density_plots}
\end{figure*}

Fig.\ \ref{density_plots} shows
VMC results for the hole density 
$$
n_h(m)\  =\ \langle 1-n_m \rangle _{\Psi_l'}~,
$$ 
in the partially projected state $|\Psi_l'\rangle$
are presented in the first row of.
The sites $m$ are distinct from the reservoir site $l$
(marked by a cross in the figure). All results shown correspond
to half filling; \textit{viz.}, $n_{\uparrow}=n_{\downarrow}=0.5$.
We choose $\Delta=1$ for the BCS states.
The vectors of periodic boundary conditions are
$\vec L_1=(L_x,L_y)$ and $\vec L_2=(-L_y,L_x)$ 
respectively, with
$L_x=37, L_y=1$; Including the reservoir site,
$L=L_x^2+L_x^2=1370$ sites. 
In the figure, white/black correspond to high/low values of 
$n_h(m)$, which
is scaled by a logarithmic gray scale varying 
in the range $-8.5 < \log n_h(m) < -6$. 
Thus, the same gray
represents the same value in all the three cases shown.

For the Fermi sea, we see that the hole is distributed 
more uniformly than the other cases even though the diagonal 
direction has a larger probability of being occupied by a hole.
The $s$-wave shows a checker-board pattern.
The $d$-wave has a quasi checker-board pattern 
where only one of four sites is black, and the 
hole tends to be near the reservoir site.
The VMC results for the projected BCS wave functions are strikingly
different in that the hole density
is \textit{not} uniform. On the other hand, the 
Gutzwiller approximation would be exact, if all states in the 
Hilbert space contribute equally to the wave function. That would
correspond to a uniform density of holes. Clearly, the Gutzwiller
approximation has to be extended to treat projected superconducting
wave functions. This
is in agreement with our previous considerations, where
we found that the
functional form of $X$ (Eq.~\ref{X_Gutzwiller}, derived using Gutzwiller
approximation) agrees with the VMC calculations only for the
projected Fermi sea, but not for BCS states (see Fig.\ \ref{d_Gutz_doping} and
Fig.\ref{d_BCS_doping}).

To further investigate the effects of Gutzwiller projection,
we also plot (second row of Fig.\ \ref{density_plots})
the correlation function
\begin{eqnarray} \label{d_h}
d_h^{(0)}(m) & = &
 \langle
n_{l\uparrow}n_{l\downarrow} (1-n_{m\uparrow})(1-n_{m\downarrow})
\rangle
\\ \nonumber
& -&
 \langle n_{l\uparrow}n_{l\downarrow}\rangle
 \langle (1-n_{m\uparrow})(1-n_{m\downarrow})\rangle
\end{eqnarray}
in systems {\it without} the Gutzwiller projection. This
correlations function between a hole at site $m$ and a doubly
occupied site at $l$ corresponds to 
the quantity $n_h(m)$ for the partially projected
wave function close to half filling. This is because, 
in the latter case, the unprojected site is doubly occupied.
Note that translation invariance implies that
the second term in (\ref{d_h}) does not
depend on the site indices $l$ and
$m$, and is a constant factor. Then, using Wick's theorem,
the correlation function $d_h^{(0)}(m)$ 
is reduced to a function of
$\langle c^\dagger_{i,\uparrow} c_{j,\uparrow}\rangle$ and $\langle
c_{i,\uparrow} c_{j,\downarrow}\rangle$.
The quantity
$d_h^{(0)}(m)$ can be evaluated
exactly, performing a Fourier transform (we use the
same system size and boundary conditions).
The logarithm of the correlation function
is scaled in the second row of Fig.\ \ref{density_plots}
by the gray scale varying in the range $(-22,-4)$.
Both $\langle c^\dagger_{i,\uparrow} c_{j,\uparrow}\rangle$ and $\langle
c_{i,\uparrow} c_{j,\downarrow}\rangle$ show
Friedel oscillations.

For the Fermi sea,
only $\langle c^\dagger_{i,\uparrow} c_{j,\uparrow}\rangle$ 
is finite. The nesting of the
Fermi surface by $Q=(\pi,\pi)$ then leads to the
to the checker-board pattern for the hole density
observed in Fig.\ \ref{density_plots} (second row).
 
For the $s$-wave, the Friedel oscillations of $\langle c^\dagger_{i,\uparrow}
c_{j,\uparrow}\rangle$ are similar to that of the Fermi sea while 
the oscillation
of $\langle c_{i,\uparrow} c_{j,\downarrow}\rangle$ is phase
shifted by $\pi/2$. Summing both contributions to
$d_h^{(0)}(m)$ 
the oscillations are smeared out.
In contrast, for the $d$-wave both 
$\langle c^\dagger_{i,\uparrow} c_{j,\uparrow}\rangle$ 
and $\langle c_{i,\uparrow} c_{j,\downarrow}\rangle$
oscillate in phase, leading to the oscillation observed.

Let us compare these results with those obtained after projection.
For the Fermi sea, we see clearly that the density oscillations are
suppressed by projection. This is likely because projection
reduces the discontinuity at the Fermi level, thereby
suppressing the nesting
by $Q$ and the corresponding Friedel oscillations.

The emergence of the checker-board pattern in the projected
$s$-wave suggests
that Gutzwiller projection affects
$\langle c_{i,\uparrow} c_{j,\downarrow}\rangle$
stronger than $\langle c^\dagger_{i,\uparrow} c_{j,\uparrow}\rangle$.
With only one contribution, the Friedel oscillations are no longer
smeared out and are observed.

Projection changes the pattern qualitatively for the $d$-wave too.
The observed pattern resembles approximately the function,
$
\sim\ \sin^2(x\pi/2) \sin^2(y\pi/2) 
$
(see Fig.\ \ref{density_plots}, top row), with $m=(x,y)$. This
indicates that the nodal points
at $(\pm\frac{\pi}{2},\pm\frac{\pi}{2})$ contribute
dominantly after projection.
Furthermore, in this case, the hole tends to stay near the
reservoir site. It means that only a part of the Hilbert space has a
large weight, leading to a deviation from the Gutzwiller
approximation. We believe this effect cannot be captured within
the Gutzwiller approximation without invoking off-site correlations
\cite{fn}.


\section{Discussion}

In this paper, we extended the Gutzwiller approximation 
scheme to construct normalized excitations and matrix elements 
for the projected Fermi sea. In typical calculations, 
one needs to determine matrix elements between partially 
projected Gutzwiller projected states, where double occupancies 
are projected out at all but one site $l$ (called the ``reservoir'' site).
The occupancy of the reservoir site, $n_l$
turns out to be an important quantity in the calculation of
matrix elements. Since the
wave function projects out double occupancies on all sites $m \neq l$,
it follows that the occupancy $n_m \leq 1$. But, $n_l = \{0,1,2\}$. Therefore,
our results for $n_l$ are nontrivial in that
the Gutzwiller approximation is extended to calculate the occupancy
at an \textit{unprojected} site.
We presented an analytical method to calculate such matrix 
elements and showed that the approximations
are in good agreement with results from variational
Monte Carlo (VMC) for the Fermi sea.
These results were used to construct normalized single particle 
excitations of the fully projected Fermi sea, and to calculate 
matrix elements for tunneling into the projected Fermi sea.

Single particle tunneling in projected BCS wave functions has been discussed 
recently, by Anderson and Ong \cite{pwa_tunnel}, and Randeria \emph{et al.}
\cite{randeria_tunnel}. In our calculations for tunneling into the projected
Fermi sea, we find the surprising result that the matrix 
elements for both particle and hole tunneling vanish 
as $n \rightarrow 1$ (half filling). Within our scheme, the
result follows from the behavior of the charge density in the 
vicinity of the reservoir site. In particular, for the 
projected Fermi sea the analytical result hinges on the 
expression for single occupancy of the reservoir site, 
Eq. (\ref{X_Gutzwiller}). As can be seen in Fig. 2, the 
analytical result does not agree with numerical calculations done
for projected BCS wave functions. This discrepancy underscores the 
importance of pairing correlations in the unprojected
wave functions, which are not taken into account
within the Gutzwiller approximation scheme.

There are
two ways by which electron correlations arise in the Gutzwiller scheme: 
one is through the
mean field or trial wave function
$|\Psi_0\rangle$, and the other \textit{via} the projection on the
subspace of no double occupancy, $|\Psi\rangle = P|\Psi_0\rangle$.
The latter effect, which results in the reduction in the size of the 
Hilbert space
can be described by combinatorial arguments, leading
to (\ref{X_Gutzwiller}).
As seen in Fig.\ \ref{d_Gutz_doping}, the analytical and
VMC results are in good agreement for the case
of the projected Fermi sea. We can trace this agreement back to 
the fact that the Fermi sea does not contain any
additional explicit correlations.

Consider instead $|\Psi_{\rm{BCS}}\rangle$, which contains
additional, molecular field correlations in the
unprojected wave function. Here, we may expect deviations
for quantities like the relative normalization
$X$ from the combinatorial result (\ref{X_Gutzwiller}).
Indeed, the VMC data presented in Fig.\ \ref{d_BCS_doping}
confirms this expectation. For instance, the data
show a qualitatively different dependence of $X$ on
doping, for the $s$-wave BCS states. The VMC data indicates
a possibly different limiting behavior for $X$ in the
limit $n\to1$, as indicated by the analysis of the data
as a function of inverse cluster-size, presented 
in Fig.\ \ref{size-dependence}.

For the $s$-wave BCS state, we observe a dramatic enhancement in
the double occupancy at the reservoir site for low doping, which we 
understand as a consequence of enhanced on-site pairing, 
relative to the Fermi liquid state. On the other hand,
The double occupancy of the reservoir site is
reduced for the $d$-wave, since the $d$-wave state
suppresses on-site pairing
fluctuations. The quantitative
behavior of the normalization ratio $X$ as a function
of doping, for projected superconducting states is thus a 
subtle problem 
which we hope to solve in the future.

We also studied the hole density near the reservoir site for projected
superconducting wave functions at half filling using VMC. The results
are shown in the top row of Fig.\ \ref{density_plots}.
For the projected Fermi sea,
we find that the hole density is uniform. 
However, for
the superconducting states, we find that projection induces oscillations in the 
hole density near the reservoir site. For the projected $d$-wave state,
we find that the hole density is mostly near the reservoir site. We believe
that the Gutzwiller approximation needs to be extended to treat pairing
correlations in the superconducting wave functions to understand these
results fully. This issue, along with the study of systems away from half filling
and their possible relevance to the checker-board pattern
observed in scanning tunneling microscopy of the high temperature superconductors
is left to future research.

\bigskip
We thank P.W. Anderson, N. P. Ong, and H. Yokoyama
for several discussions. N.F. is supported by the Deutsche Forschungsgemeinschaft.
V.N.M. acknowledges partial financial support from The City University of 
New York, PSC-CUNY Research Award Program.




\begin{thebibliography}{99}
\bibitem{fn}
However, we have indications that the hole is not bound 
to the reservoir site, since
the density correlation seems to decay algebraically and
because they decrease in magnitude with increasing lattice
size.
\end{thebibliography}
\end{document}